\documentclass[reprint,superscriptaddress,amsmath,amssymb,aps,prb,floatfix]{revtex4-2}
\usepackage{graphicx}
\usepackage{dcolumn}
\usepackage{bm}
\usepackage[dvipsnames]{xcolor}

\begin{document}

\title{Quantum versus thermal fluctuations in the fcc antiferromagnet: \\
alternative routes to order by disorder}

\author{R. Schick}
\affiliation{Institut Laue Langevin, 38045 Grenoble Cedex 9,France}
\affiliation{Universit\'e Grenoble Alpes, CEA, IRIG, PHELIQS, 38000 Grenoble, France}
\author{T. Ziman}
\affiliation{Institut Laue Langevin, 38045 Grenoble Cedex 9,France}
\affiliation{Universit\'e Grenoble Alpes, CNRS, LPMMC, 38000 Grenoble, France}
\author{M. E. Zhitomirsky}
\affiliation{Universit\'e Grenoble Alpes, CEA, IRIG, PHELIQS, 38000 Grenoble, France}

\date{\today}

\begin{abstract}
In frustrated magnetic systems with competing interactions fluctuations can lift 
the residual accidental degeneracy. We argue that the state selection may have 
different outcomes for quantum and thermal order by disorder. As an example, 
we consider the semiclassical Heisenberg fcc antiferromagnet with only 
the nearest-neighbor interactions. Zero-point oscillations select the type 3 
collinear antiferromagnetic state at $T=0$. Thermal fluctuations favor instead 
the type 1 antiferromagnetic structure. The opposite tendencies result in a 
finite-temperature transition between the two collinear states. Competition between 
effects of quantum and  thermal order by disorder is a general  phenomenon and is 
also realized in the $J_1$--$J_2$ square-lattice antiferromagnet at the critical 
point $J_2=\frac{1}{2}J_1$.
\end{abstract} 

\maketitle

{\it Introduction}.\ 
As counterintuitive as it may seem, fluctuations are not always destructive of 
order, but can actually stabilize broken symmetry states. One prominent example 
is the isotropic-nematic transition in liquid crystals consisting of hard-rod 
molecules that, according to Onsager, can be driven entirely by entropy gain in 
the ordered state \cite{Onsager49}. In high-energy physics, vacuum fluctuations 
are the hallmark of the Coleman-Weinberg mechanism of spontaneous symmetry breakdown 
in the electrodynamics of massless scalar mesons \cite{Coleman73}. The concept of 
fluctuation-induced ordering has also gained a lot of attention in the field of 
magnetism. Frustrated magnets with competing interactions often exhibit accidental 
degeneracy between classical or mean-field ground states that is not dictated by 
symmetry \cite{Shender95,Moessner01,HFM10}. Their low-temperature behavior and 
the ultimate ground state selection is sensitive to weak residual interactions 
but can also be determined solely by fluctuations. In the past, several authors 
have independently shown that accidental degeneracy in frustrated spin models 
can be lifted either by quantum or by thermal fluctuations 
\cite{Tessman54,Belorizky76,Villain80,Shender82,Rastelli83,Kawamura84}.
Nowadays, such a fluctuation mechanism is commonly referred to as the effect 
of order by disorder
\cite{Rastelli88,Henley89,Chandra90,Chubukov91,Chalker92,Sheng92,Yildirim96,Moessner98,Zhito00,%
Bergman07,Bernier08,Zhito12,Lee14,Chernysh14,Javan15,Jackeli15,Rouso15,Rau16,Danu16,Rau18}.

The early theoretical works have typically found selection of the `most collinear' 
states for both quantum ($T=0$) and classical ($T>0$) versions of the same 
frustrated spin model, supporting the perception that the two types of fluctuations 
play a similar role in the state selection process. Being correct for many 
weakly frustrated magnets, this assertion is, however, not guaranteed in general, 
see examples in \cite{Bernier08,Chernysh14,Danu16}.

A further limitation of the current picture of thermal order by disorder 
is that it is mostly based on studies of classical spins. This is, in part,
because only the classical Monte Carlo algorithms have proved efficient for 
frustrated models. The role of thermal fluctuations in quantum spin 
models attracted much less attention. 
The available works on this problem \cite{Rastelli88,Sheng92,Lee14} have 
documented a distinct role of thermal fluctuations for a few specific
spin models, but generality of the obtained results remain unclear.

The difference between quantum and thermal order by disorder is easily 
recognized by considering two standard expressions. At $T=0$, the zero-point 
(vacuum) energy is given by
\begin{equation}
E_0  = \frac{1}{2} \sum_{\bf k} \epsilon_{\bf k} \ ,
\label{E0} 
\end{equation}
where $\epsilon_{\bf k}$ are energies of bosonic magnon modes. At the same time, 
their free energy is ($k_B=1$)
\begin{equation}
\Delta F = T \sum_{\bf k} \ln \Bigl(1 - e^{-\epsilon_{\bf k}/T}\Bigr)\,.
\label{DFT}
\end{equation}
Both types of fluctuations favor states with soft excitations but the `softness' 
criterion appears to be different in each case. Minimization of the zero-point energy (\ref{E0}) 
picks the states with the smallest average magnon energy. Thermal fluctuations (\ref{DFT})  
instead select states with the largest density of low-energy excitations 
$\epsilon_{\bf k}\sim T$.  In frustrated magnets the low-energy excitations include the pristine 
Goldstone modes determined by the broken symmetry and the so-called pseudo-Goldstone modes. 
The latter appear due to an accidental degeneracy of the classical ground states and 
have a distinct structure for each of the ground states. Since the two selection mechanisms
rely on magnons with different energies, their outcome can also vary.

The thermal effects vanish as $T\to 0$, hence, a natural question is whether 
the thermal contribution (\ref{DFT}) can overcome the zero-temperature splitting (\ref{E0}). 
In our paper we demonstrate that the semiclassical Heisenberg antiferromagnet (AFM) 
on a face-centered cubic (fcc) lattice with nearest-neighbor interactions exhibits 
a finite-temperature transition within magnetically ordered state determined by 
competition between the two fluctuation mechanisms. A competition between thermal and quantum
order by disorder effects is also predicted for the frustrated square-lattice antiferromagnet
showing that such a behavior is ubiquitous among frustrated models. 

{\it FCC antiferromagnet, $T=0$}.\ The Heisenberg AFM on an fcc lattice is one of the oldest 
frustrated spin models \cite{Anderson50,Ziman53,Haar62,Lines63,Yamamoto72,Swendsen73}. 
It keeps attracting significant interest because of numerous experimental realizations
\cite{Seehra88,Matsuura03,Goodwin07,Balagurov16,Aczel16,Chatterji19,Khan19,Revelli19}. 
We consider the Heisenberg model with spins of length $S$ and the nearest-neighbor 
exchanges of strength $J$:
\begin{equation}
{\cal H} = J\sum_{\langle ij\rangle} {\bf S}_i\cdot{\bf S}_j \ .
\label{Hnn}
\end{equation}
Each spin couples to 12 nearest neighbors located at $(0,\pm a/2,\pm a/2)$, 
$(\pm a/2, 0,\pm a/2)$, $(\pm a/2,\pm a/2,0)$, where $a$ is the linear size 
of a cubic cell.

\begin{figure}[tb]
\centering
\includegraphics[width=0.48\linewidth]{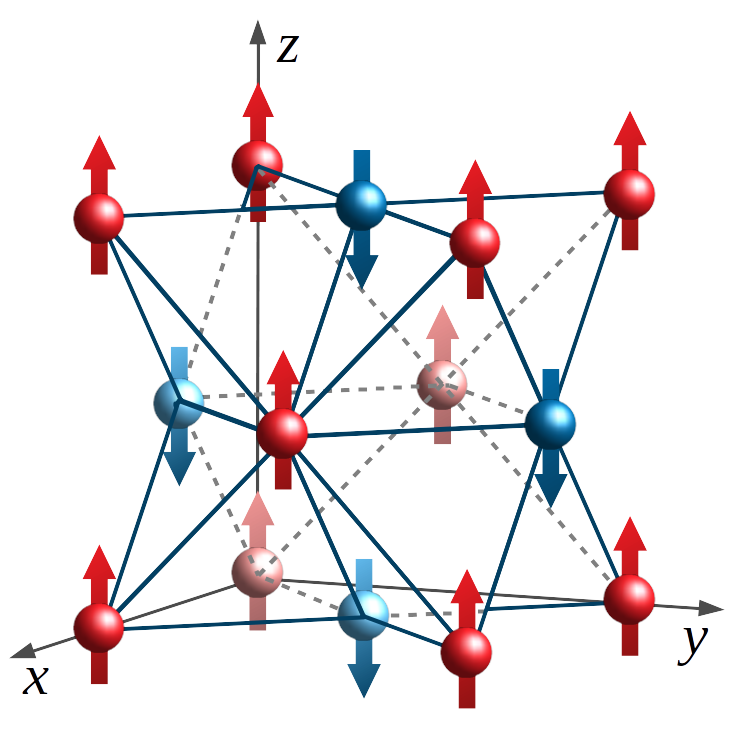} \hspace{1mm}
\includegraphics[width=0.48\linewidth]{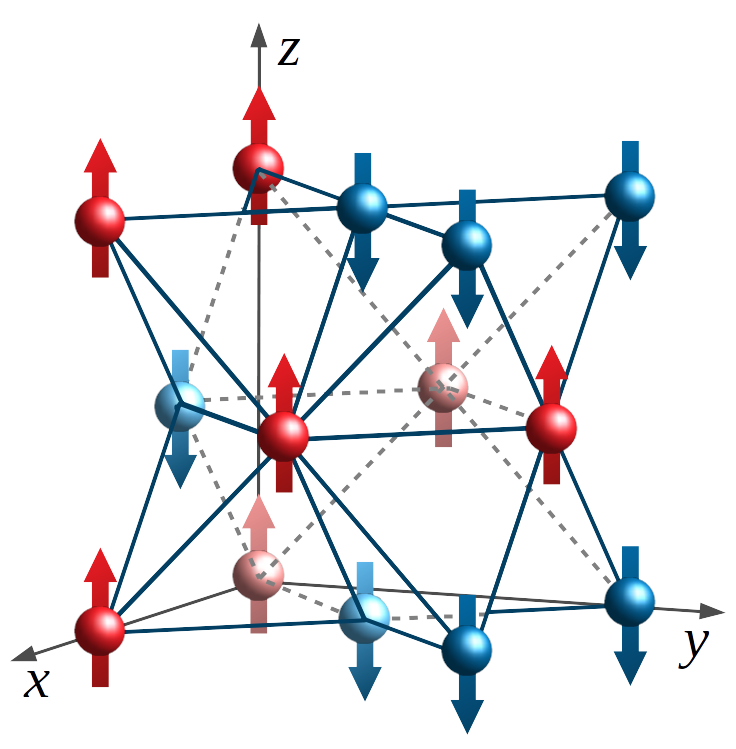} 
\caption{Type 1 (left) and  type 3 (right) antiferromagnetic structures on 
an fcc lattice.}
\label{fig:TypeI_III}
\end{figure}

The lowest-energy classical states of (\ref{Hnn}) are coplanar spin spirals with 
the propagation vectors belonging to the line
\begin{equation}
{\bf Q}_s = \frac{2\pi}{a}(1,q,0)
\label{Qcl}
\end{equation}
and other equivalent directions in the cubic Brillouin zone. Their classical energy 
$E_{cl} = -2JS^2$ does not depend on the pitch parameter $q$. For each degeneracy 
line there are two special commensurate wave vectors 
\begin{equation}
{\bf Q}_1 = \frac{2\pi}{a}(1,0,0) \quad  \textrm{and} \quad 
{\bf Q}_3 = \frac{2\pi}{a}\Bigl(1,{\frac{1}{2}},0\Bigr) 
\label{Q1Q3}
\end{equation}
that accommodate collinear states called respectively the type 1 (AF1) and type 3
antiferromagnetic (AF3) structures, see Fig.~\ref{fig:TypeI_III}. The two collinear states 
become unique ground states in the presence of a weak second-neighbor 
exchange either of FM (type 1) or AFM (type 3) sign \cite{Yamamoto72}. 
This makes them natural candidates for the order by disorder selection in the 
nearest-neighbor case \cite{Henley87}. 

Significant efforts were previously devoted to investigation of thermal order 
by disorder for the classical fcc antiferromagnet 
\cite{Henley87,Fernandez83,Minor88,Diep89,Alonco96,Gvozdikova05}. The large-scale 
Monte Carlo simulations clearly demonstrated presence of the AF1 state below 
the first-order transition at $T_c\approx 0.446JS^2$ \cite{Gvozdikova05}.
Surprisingly, the quantum selection for the Heisenberg fcc antiferromagnet (\ref{Hnn})
was not addressed in detail apart from one early work \cite{Haar62}, which came, as is 
shown below, to an incorrect conclusion.  The main focus of more recent theoretical studies 
\cite{Oguchi85,Yildirim98,Fishman98,Ader01,Ignatenko08,Datta12,Ishizuka15,Sinkovicz16,%
Batalov16,Li17,Singh17}
was on the effect of further-neighbor exchanges, anisotropies etc.

We consider semiclassical spins $S\gg 1$ and use the linear spin-wave theory (LSWT)
to study the ground state selection by quantum fluctuations. Spin operators are bosonized 
via the Holstein-Primakoff transformation applied in the rotating local frame and only 
quadratic terms in boson operators are kept. The quadratic form is diagonalized by the 
Bogolyubov transformation allowing to compute the magnon dispersion $\epsilon_{\bf k}$ 
and the ground-state energy per spin 
\begin{equation}
E_{\textrm{g.s.}} = -2JS(S + 1) + \frac{1}{2}\sum_{\bf k}\epsilon_{\bf k}  \ ,
\label{Egs}
\end{equation}
where summation is taken over the first Brillouin zone. All steps are completely 
standard, see, e.g., \cite{Zhito96}, and below we present only the final expressions.

\begin{figure}[tb]
\centering
\includegraphics[width=0.65\columnwidth]{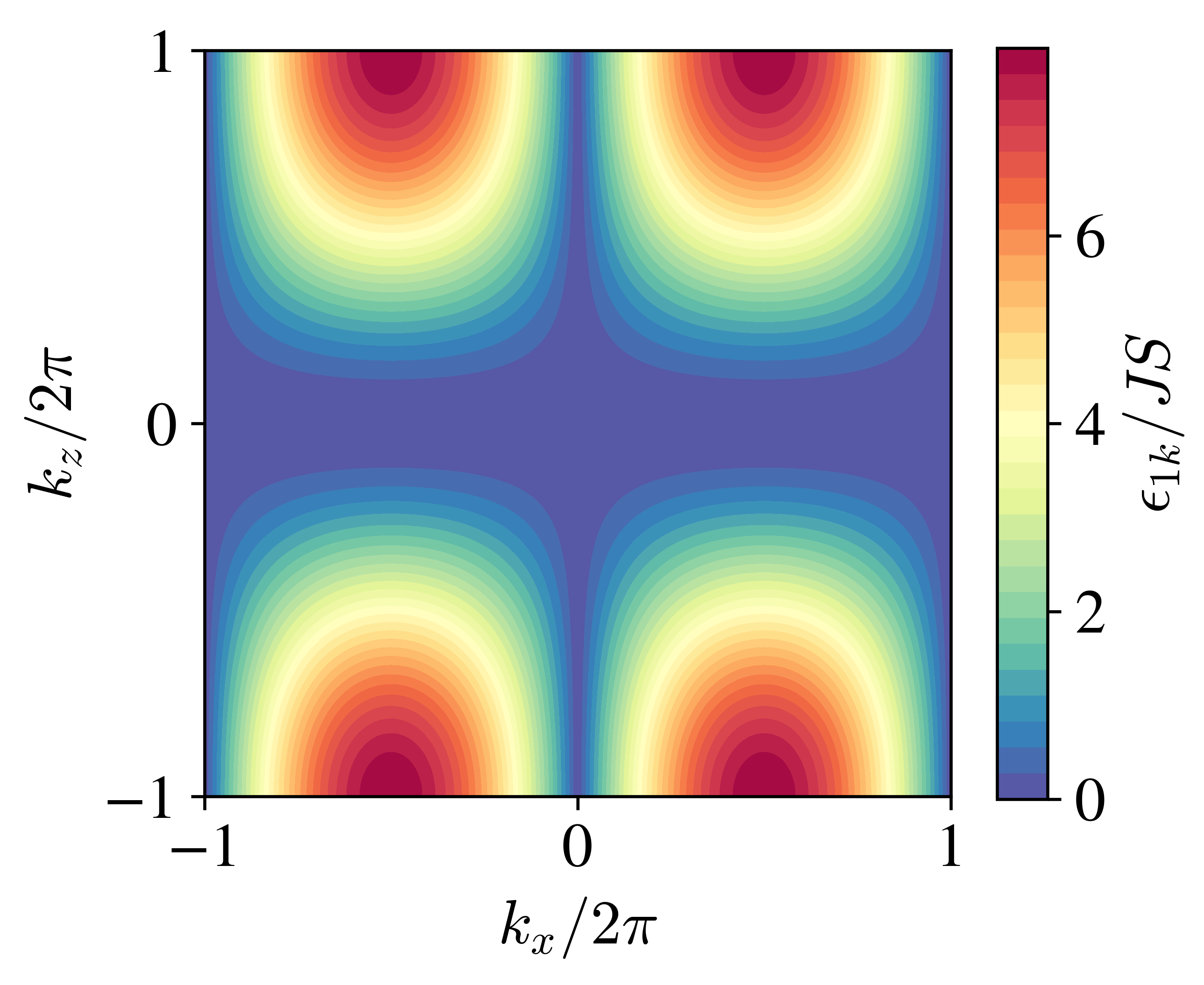}
\\[0mm]
\includegraphics[width=0.65\columnwidth]{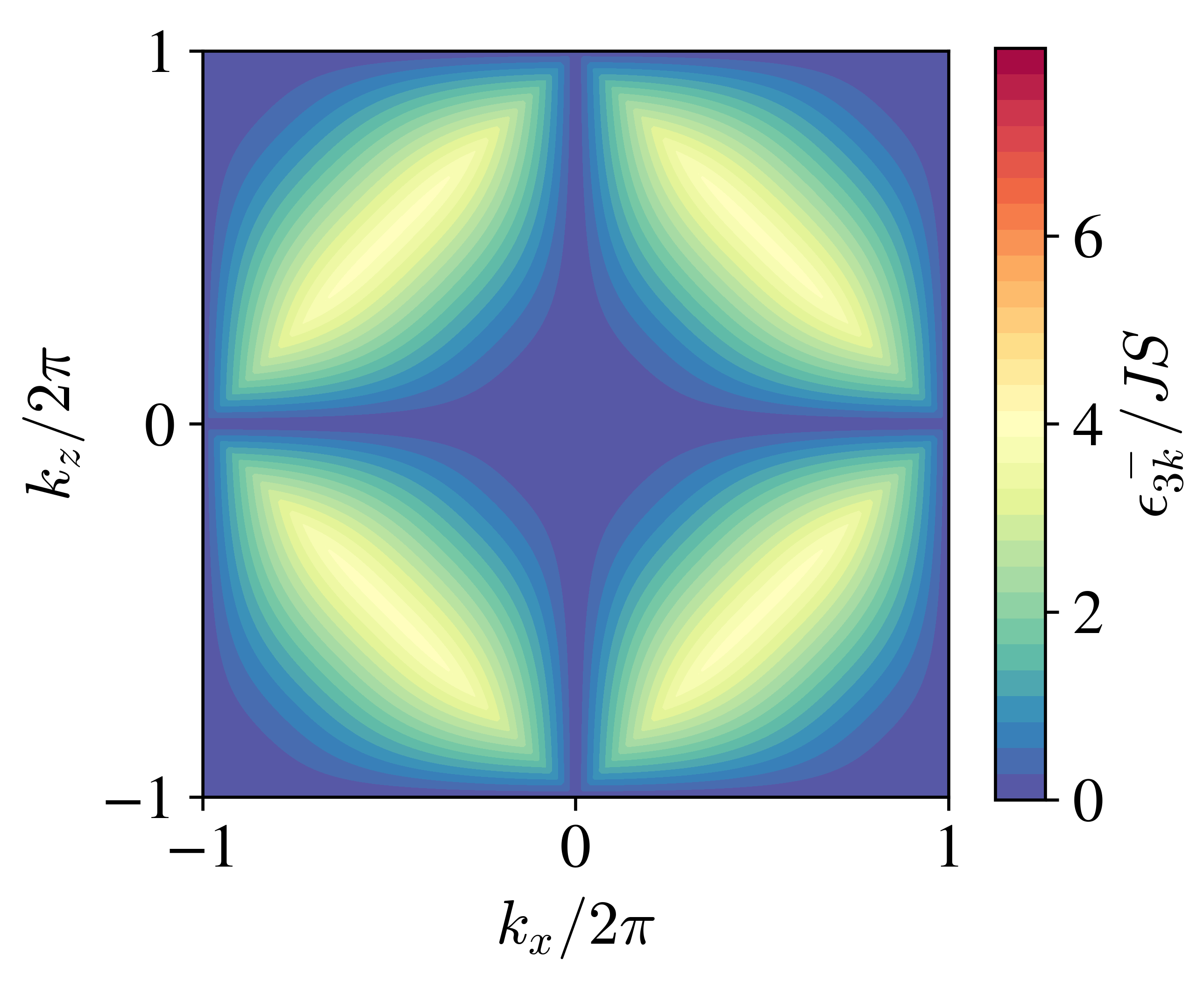}
\caption{Color intensity map for the magnon dispersion in 
two collinear antiferromagnetic states with fixed $k_y = 2\pi/a$: 
AF1 (top panel) and  AF3, lower branch, 
(bottom panel).}
\label{fig:Ekmap}
\end{figure}

For an arbitrary spin spiral on the degeneracy line (\ref{Qcl}) the magnon energy is 
\begin{eqnarray}
\epsilon_{s\bf k} & = & 4JS\,\bigl[1+c_xc_y+c_xc_z+c_yc_z\bigr]^{1/2}
 \nonumber   \\
& & \mbox{} \times
\bigl[1-c_xc_z+c_y(c_z-c_x)\cos(\pi q)\bigr]^{1/2}\ .
\label{eks} 
\end{eqnarray}
To simplify formulas we define
$c_{\alpha} = \cos(k_{\alpha}a/2)$, $s_{\alpha} = \sin(k_{\alpha}a/2)$ for 
$\alpha =x,y,z$. The excitation spectrum for the AF1 state 
is obtained by taking $q\to 0$ in the above equation:
\begin{equation}
\epsilon_{1\bf k} = 4JS\sqrt{s_x^2 (c_y + c_z)^2 + s_y^2 s_z^2 } \  .  
\label{ek1}
\end{equation} 
For  $q = 1/2$, Eq.~(\ref{eks}) describes magnons in the noncollinear spiral state 
with the propagation vector ${\bf Q}_3$ and the $90^\circ$ angle between 
neighboring spins. The spectrum of the collinear AF3 state cannot be described
in a simple rotating basis. One has to include two sites in the unit cell 
and introduce two types of bosons. Accordingly, there are two magnon modes 
for each wave vector
\begin{eqnarray}
\Bigl(\frac{\epsilon_{3\bf k}^{\pm}}{4JS}\Bigr)^2 & = &
1 - c_x^2 c_z^2 \pm \Bigl[c_y^2 (c_x + c_z)^2 (1 - c_xc_z)^2  \nonumber \\
& + & s_y^4(c_x^2 - c_z^2)^2  +  s_x^2 s_y^2 s_z^2 (c_x - c_z)^2 \Bigr]^{1/2} ,
\label{ek3}
\end{eqnarray} 
which both contribute to the zero-point energy, but the momentum summation is now 
performed over a half of the Brillouin zone. The dispersion relation (\ref{ek3}) was 
previously derived by Swendsen \cite{Swendsen73}. However, the early work by ter Haar 
and Lines \cite{Haar62,Lines63} gave $\epsilon_{\bf k}$ equivalent to Eq.~(\ref{eks}) 
with $q=1/2$, which applies to the ${\bf Q}_3$ spiral rather than to the collinear AF3 
state \cite{remark}. The magnon dispersion in the two collinear states is illustrated in 
Fig.~\ref{fig:Ekmap}. Apart from the normal Goldstone modes at the momenta 
${\bf k}=0$ and ${\bf k}={\bf Q}_1$ or ${\bf Q}_3$ the excitation spectra contain 
the line nodes that appear due to the classical degeneracy. 

\begin{figure}[b]
\centering
\includegraphics[width = 0.9\columnwidth]{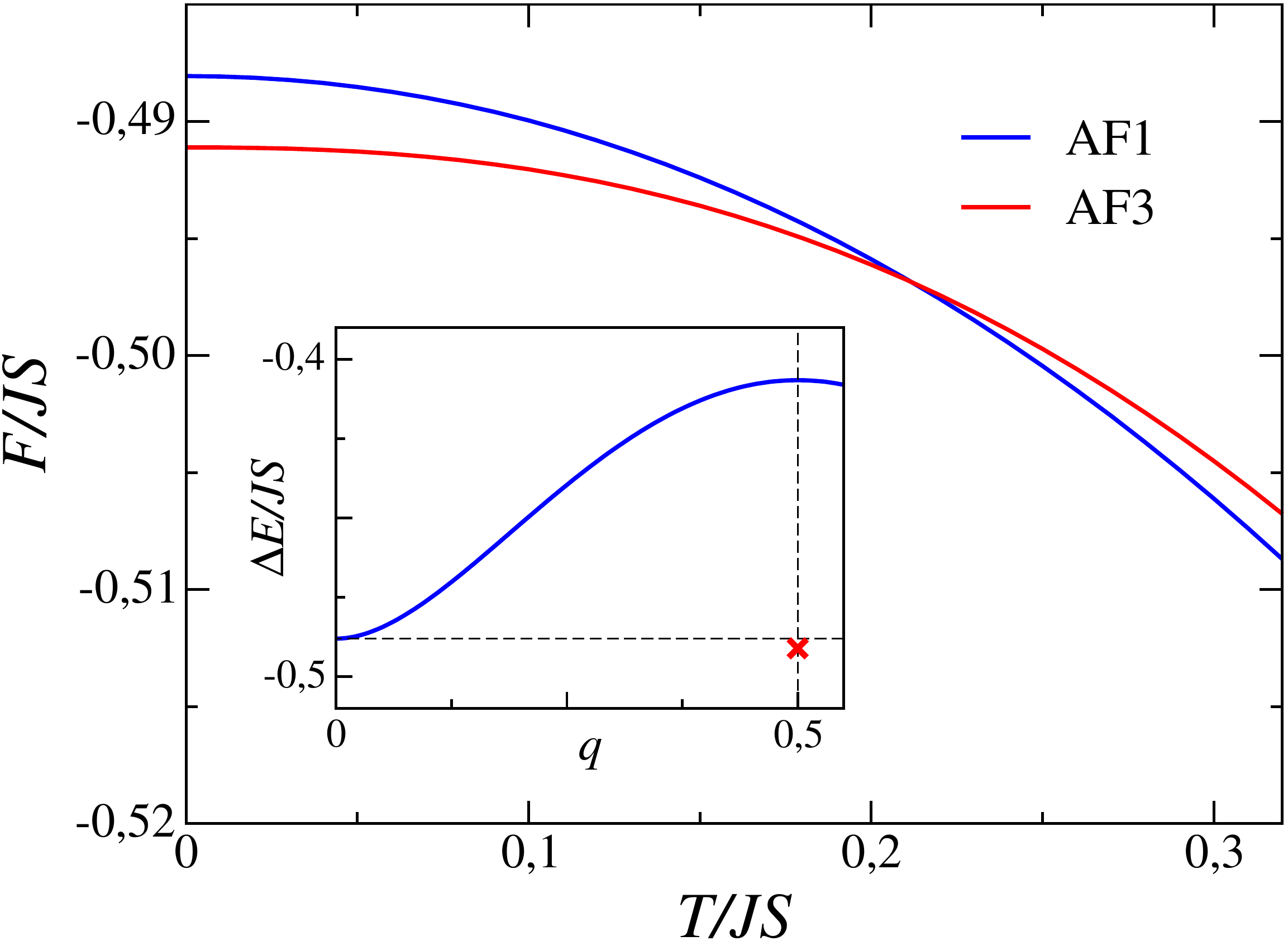}
\caption{Main panel: temperature dependence of the free energy in the AF1 and 
AF3 states.  Inset: the quantum correction $\Delta E$ to 
the ground-state energy for the degenerate classical spirals versus a pitch $q$ 
(full line). $\Delta E$ for the AF3 state is marked with a cross.
}
\label{fig:FvsT}
\end{figure}

The zero-point energy for degenerate classical ground states has been computed 
numerically using the magnon spectra (\ref{eks})--(\ref{ek3}). 
Results quoted below contain all significant digits.
Combining $E_0$ with the state-independent negative shift, Eq.~(\ref{Egs}),
we obtain for the ground-state energy of the AF1 state 
\begin{equation} 
E_{\rm g.s.}^{(1)} = E_{cl} \Bigl( 1 +  \frac{0.488056}{2S}\Bigr)\,,
\label{Egs1}
\end{equation} 
where $E_{cl} = -2JS^2$. The first three digits of the $1/S$  correction 
agree with the result of Ref.~\cite{Oguchi85}. For the AF3 structure the integration yields
\begin{equation} 
E_{\rm g.s.}^{(3)}  = E_{cl} \Bigl( 1 +  \frac{0.491106}{2S}\Bigr)\,,
\label{Egs3}
\end{equation} 
which is lower than $E_{\rm g.s.}^{(1)}$. The inset of Fig.~\ref{fig:FvsT} 
shows the quantum energy correction $\Delta E = E^{(s)}_{\rm g.s.} - E_{cl}$
for the spin spirals (\ref{Qcl}), which is always above  $\Delta E^{(1)}$
\cite{remark}. Thus, at variance with \cite{Haar62} we find that the LSWT
gives the lowest energy for the AF3 state albeit with a rather small energy difference  
$\Delta E_0 \approx 0.003JS$.  This conclusion is  further supported by
 the numerical exact-diagonalization study of the spin-1/2 model \cite{Lefmann01},
 which found enhanced spin-spin correlations at the wave vector ${\bf Q}_3$ in comparison to 
${\bf Q}_1$.  Still, a small size of the employed cluster ($N=32$) prevents from making
any definite statement about the state selection in the $S=1/2$ case.
Therefore, it will be interesting to check how  the higher-order 
spin-wave corrections modify the ground-state energies  of the AF1 and AF3 spin structures.

We conclude the $T=0$ case with results for the ordered moments. 
Due to the additional pseudo-Goldstone modes  the spin reduction 
$\Delta S = S-\langle S\rangle$  is substantial for both states:
\begin{equation}
\Delta S_1 = 0.33875 \ , \quad
\Delta S_3 = 0.36630 \ .
\label{dS}
\end{equation} 
In agreement with the fluctuation mechanism, the lowest energy  state exhibits 
a larger spin reduction.

{\it FCC antiferromagnet, finite $T$}.\ 
The free energy (\ref{DFT}) has been computed in the low-temperature region $T\ll JS^2$ 
using the bare magnon spectra. We normalize 
temperature to $JS$ and drop the classical energy, which leaves the same 
$JS$-scaling for both contributions (\ref{E0}) and (\ref{DFT}).  
Figure~\ref{fig:FvsT} shows the total free energy $F=E_0+\Delta F$ for the two 
collinear states. Remarkably, curves cross at $T^* \approx 0.21JS$ 
indicating the first-order transition into the AF1 state
above $T^*$.  This is an interesting example of competition between thermal and 
quantum order by disorder. Clearly, the transition is possible because of 
a small initial difference in the zero point energies of the competing 
states.

To further understand the thermal vs.\ quantum competition we derive analytically  
the low-temperature asymptotes for $\Delta F(T)$ in the two states. Let us begin 
with the AF1 structure. The energy (\ref{ek1}) has two types of line nodes:  
(i) $L_x$-line ${\bf k} =(k_x,2\pi/a,0)$ and (ii) $L_z$-line ${\bf k} =(0,2\pi/a,k_z)$. 
All other zero-energy excitations fall in one of the above categories. Difference 
between $L_x$ and $L_z$ nodes is prominent already from Fig.~\ref{fig:TypeI_III}. 
The small momentum expansion around the $L_x$-line gives
\begin{equation}
\epsilon_{1\bf k} \approx JSa^2 \sqrt{k_y^2k_z^2 + {\textstyle\frac{1}{4}}s_x^2(k_y^2-k_z^2)^2}
\simeq k_\perp^2 \ ,
\label{e1kA}
\end{equation}
whereas for the $L_z$ line  $\epsilon_{1\bf k}\simeq k_\perp$. The softer $L_x$ 
magnons dominate at low temperatures and yield the power law asymptote
\begin{equation} 
\Delta F_1 \propto -T^2 
\label{DF1}
\end{equation}
and $C\propto T$  behavior of the specific heat. These
should be contrasted with a much weaker thermal effect $\Delta F \propto -T^4$ 
($C\propto T^3$) in nonfrustrated 3D antiferromagnets.
 
The magnon dispersion for the AF3 state also has line nodes, but they have a 
linear dispersion and play only a secondary role. The dominant contribution 
into $\Delta F_3$ comes from the crossing points of such lines, see
Fig.~\ref{fig:TypeI_III}. In the vicinity of the crossing point $(0,2\pi/a,0)$ 
the lowest magnon branch (\ref{ek3}) is anomalously soft
\begin{equation} 
\epsilon_{3\bf k}^- \approx \frac{JSa^2}{\sqrt{2}} \sqrt{k_y^2(k_x^2 + k_z^2)+ 
{\textstyle\frac{1}{16}}\,k_x^2 k_z^2(k_x^2 + k_z^2)}\ ,
\label{ek3A}
\end{equation} 
where both the leading and the subleading terms are necessary
for deriving the correct asymptote. A lengthy analytic calculation
gives in this case 
\begin{equation}
\Delta F_3 \propto -T^{7/3} \ .
\label{DF3}
\end{equation}
The two power laws  (\ref{DF1}) and (\ref{DF3}) fully agree with the numerical 
results in Fig.~\ref{fig:FvsT}. Thus, the different thermal response in the two states 
is determined by a different structure of the pseudo-Goldstone modes.

Generally, the higher-order $1/S$ corrections renormalize the harmonic spectrum 
and induce a small quantum gap $\Delta_g =O(1/S)$ for the pseudo-Goldstone modes.
Such calculations have been carried out before in a few simple cases
\cite{Yildirim98,Khaliulin01,Holt11,Rau18}. The normal 3D behavior
$\Delta F \propto -T^4$ is recovered  at ultra low temperatures 
$T\ll\Delta_g$ leaving almost intact the $T$-dependences (\ref{DF1}) and (\ref{DF3}) in the
intermediate regime $\Delta_g\ll T\alt JS$. Note, that the nonlinear effects also remove 
the divergent finite-$T$ contribution to the sublattice magnetization, which was
regarded as an indication of the absence of a long-range order 
at finite temperatures in the fcc antiferromagnet \cite{Ziman53,Haar62}.

The predicted transition at $T^* \approx 0.21JS$ occurs away from the classical 
regime $T\sim JS^2$, where a large-$S$ model behaves essentially as a classical 
spin system. Hence, the thermal order by disorder effect discussed here is not directly related 
to the similar selection in the classical model. Indeed, the harmonic excitation 
spectra are identical for the classical AF1 and AF3 states and the thermal 
selection relies on the nonlinear processes \cite{Gvozdikova05}.

{\it Frustrated square-lattice antiferromagnet.}\ We now briefly consider the Heisenberg 
$J_1$--$J_2$ antiferromagnet on a square lattice 
(FSAFM), for review see \cite{Schmidt17}. Depending on the ratio of two exchanges the model 
has the following classical ground states: the N\'eel state with ${\bf Q}=(\pi,\pi)$ 
for $J_2/J_1\leq 1/2$ and the stripe state with ${\bf Q}=(\pi,0)$ or ${\bf Q}=(0,\pi)$ 
for $J_2/J_1\geq 1/2$. We focus on the critical point $J_2/J_1=1/2$, where the classical 
degeneracy of FSAFM is reminiscent of the fcc antiferromagnet. 
Apart from the two degenerate collinear states there is an infinite number of spin 
spirals of equal energy with ${\bf Q}=(\pi,q)$ and ${\bf Q}=(q,\pi)$. 

We perform the LSWT calculations at $J_2/J_1=1/2$ in the two 
collinear states of FSAFM \cite{Bruder92}, and obtain
the magnon dispersion in the N\'eel state as
\begin{equation} 
\epsilon_{\bf k} = 2J_1S |\sin k_x| |\sin k_y| \ ,
\label{ekN}
\end{equation}
whereas  in the stripe phase [${\bf Q}=(\pi,0)$] the dispersion is
\begin{equation} 
\epsilon_{\bf k} = 2J_1S |\sin k_x| (1+\cos k_y) \ . 
\label{ekS}
\end{equation}
The excitation spectra possess the line nodes in accordance with the classical
degeneracy, but the asymptotic behavior of  $\epsilon_{\bf k}$ in their vicinity 
is markedly different between the two states. An elementary integration yields 
for the zero-point energy 
\begin{equation} 
E_0^{\rm Neel} = \frac{4}{\pi^2} J_1S < E_0^{\rm stripe} = \frac{2}{\pi} J_1S \ ,
\label{E02D}
\end{equation}
with the N\'eel state having a lower energy because of a more narrow magnon bandwidth. 
Thus, at $J_2= \frac{1}{2}J_1$, the quantum fluctuations stabilize the N\'eel state
in full agreement with the numerical evaluation of two spin-wave contributions
 \cite{Jackeli04}.

At low temperatures the leading contributions to the free energy are 
also straightforwardly computed as
\begin{equation} 
\Delta F^{\rm Neel} = - T^2\ln T\ , \quad  \Delta F^{\rm stripe} = - T^{3/2} \ .
\label{DFT2D}
\end{equation}
The stripe phase is favored by the thermal fluctuations due to its softer 
pseudo-Goldstone modes ($\epsilon_{\bf k}\sim k_y^2$).
Hence, FSAFM provides another example of the thermal-quantum competition, though
a finite-temperature transition (crossover) between the two competing state is 
precluded  due to a fairly large energy difference $\Delta E_0 \approx 0.23J_1S$ 
at zero $T$.  It will be interesting to check how the interlayer
coupling present  in possible experimental realizations of FSAFM affect this competition.

{\it Conclusions.}\ 
The difficulties in the spin-wave theory for the fcc antiferromagnet were recognized 
more than half a century ago \cite{Ziman53}, but not really resolved. By making a full linear 
spin-wave analysis of the Heisenberg model we have elucidated a number of interesting 
points: we find a qualitative difference between the effects of quantum corrections and 
thermal excitations and predict a phase transition between two collinear states each 
favored by specific type of fluctuations. We relate the differences to the structure of 
zero-frequency (pseudo-Goldstone) magnons that are responsible for the anomalous power laws  
in the free energy as a function of temperature. These should be visible in the temperature 
dependence  of the specific heat and may serve as an experimental hint of order by disorder 
in real materials. A similar transition due to competing order by disorder effects
has been found for the Heisenberg-Kitaev model on a hyper-honeycomb lattice \cite{Lee14},
though the role played by pseudo-Goldstone modes was not investigated.

The obtained results are valid for the fcc antiferromagnets with large spins $S\gg 1$. 
For small spins, the higher-order quantum corrections may become important and it is 
necessary to investigate their effect both analytically and numerically.

Competition between quantum and thermal order by disorder must be ubiquitous 
among frustrated magnets. The primary candidates are spin systems with degeneracy 
along lines in the momentum space similar to the fcc and FSAFM models considered here. 
Such spiral degeneracy naturally appears from the frustrating further-neighbor Heisenberg 
exchanges \cite{Rastelli79,Niggemann20}. It can also arise from anisotropic 
nearest-neighbor interactions on geometrically frustrated lattices \cite{Zhito12,Jackeli15,Li17}. 
We hope that presented results will stimulate further interest in the role
of thermal fluctuations in quantum frustrated models.

{\it Acknowledgments.}\ We thank  Y. Iqbal, P. Grigoriev, and J. Villain for useful discussion. 
The work of MEZ was supported by the ANR grants Matadire and Fragments.

\end{document}